\documentclass[aps,prl,reprint,groupedaddress]{revtex4-1}
\usepackage{graphicx} 
\usepackage{dcolumn} 
\usepackage{bm}
\usepackage{hyperref}

\begin{document}

\title{Nuclear in-medium and isospin effects on subthreshold kaon production in heavy-ion collisions}

\author{Zhao-Qing Feng}
\email{fengzhq@impcas.ac.cn}
\affiliation{Institute of Modern Physics, Chinese Academy of Sciences, Lanzhou 730000, People's Republic of China}

\date{\today}

\begin{abstract}
Subthreshold kaon (K$^{0}$ and K$^{+}$) production in neutron-rich nuclear reactions to probe the kaon-nucleon interaction in nuclear medium and to extract the isospin dependent part of the nuclear equation of state at high-baryon densities, is investigated within an isospin and momentum dependent transport model. A repulsive kaon-nucleon potential is implemented in the model through fitting the flow data and inclusive spectra in heavy-ion collisions, which enhances the energetic kaon emission squeezed out in the reaction zone and leads to a variation of the high-momentum spectrum of the K$^{0}$/K$^{+}$ yields. It is found that the stiffness of nuclear symmetry energy plays a significant role on the isospin ratio with decreasing the incident energy and a hard symmetry energy has a larger value of the K$^{0}$/K$^{+}$ ratio in the domain of subthreshold energies.
\begin{description}
\item[PACS number(s)]
21.65.Ef, 24.10.Lx, 25.75.-q
\end{description}
\end{abstract}

\maketitle

Kaon production in relativistic heavy-ion collisions has been investigated as a useful tool to constrain the high-density information of isospin symmetric nuclear equation of state (EoS) both experimentally \cite{La99,Fo07,Me07} and theoretically \cite{Ai85,Li97,Fu01,Ha06,Fe10}. Kaons ($K^{0}$ and $K^{+}$) as a probe of EoS are produced in the high-density domain without subsequent reabsorption in nuclear medium. The available experimental data already favored a soft EoS at high baryon densities. The $K^{0}/K^{+}$ ratio was proposed as a sensitive probe to extract the high-density behavior of the nuclear symmetry energy (isospin asymmetric part of EoS) \cite{Fe06,Pr10}, which is poorly known up to now but has an important application in astrophysics, such as the structure of neutron star, the cooling of protoneutron stars, the nucleosynthesis during supernova explosion of massive stars etc \cite{St05}. Produced kaons in heavy-ion collisions can be easily deviated by surrounding nucleons in the dynamical evolutions. Consequently, the spectrum of the $K^{0}/K^{+}$ yields is to be modified and the constraint of the density dependence of symmetry energy from heavy-ion collisions is also influenced.

In this work, kaon dynamics in heavy-ion collisions is investigated with an isospin and momentum dependent transport model (Lanzhou quantum molecular dynamics (LQMD)), in which strangeness production is contributed from channels of baryon-baryon and pion-baryon collisions \cite{Fe10}. The model has been developed to treat the nuclear dynamics at near Coulomb barrier energies and also to describe the capture of two heavy colliding nuclides to form a superheavy nucleus \cite{Fe05}. Further improvements of the LQMD model have been performed in order to investigate the dynamics of pion and strangeness productions in heavy-ion collisions and to extract the information of isospin asymmetric EoS at supra-saturation densities \cite{Fe10,Fe09}. The momentum dependence of the symmetry potential was also implemented in the model, which results in an isospin splitting of proton and neutron effective mass in nuclear medium \cite{Fe11}. We have included the resonances ($\Delta$(1232), N*(1440), N*(1535)), hyperons ($\Lambda$, $\Sigma$) and mesons ($\pi$, $K$, $\eta$) in hadron-hadron collisions and the decays of resonances for treating heavy-ion collisions in the region of 1A GeV energies.

In the LQMD model, the time evolutions of the baryons (nucleons and resonances), hyperons and mesons in reaction system under a self-consistently generated mean-field are governed by Hamilton's equations of motion, which read as
\begin{eqnarray}
\dot{\mathbf{p}}_{i}=-\frac{\partial H}{\partial\mathbf{r}_{i}},
\quad \dot{\mathbf{r}}_{i}=\frac{\partial H}{\partial\mathbf{p}_{i}}.
\end{eqnarray}
The Hamiltonian of baryons consists of the relativistic energy, the effective interaction potential and the momentum dependent part as follows:
\begin{equation}
H_{B}=\sum_{i}\sqrt{\textbf{p}_{i}^{2}+m_{i}^{2}}+U_{int}+U_{mom}.
\end{equation}
Here the $\textbf{p}_{i}$ and $m_{i}$ represent the momentum and the mass of the baryons. The effective interaction potential is composed of the Coulomb interaction and the local potential \cite{Fe11}. The local interaction potential is derived from the Skyrme energy-density functional as the form of
$U_{loc}=\int V_{loc}(\rho(\mathbf{r}))d\mathbf{r}$. The energy-density functional reads
\begin{eqnarray}
V_{loc}(\rho)=&& \frac{\alpha}{2}\frac{\rho^{2}}{\rho_{0}} +
\frac{\beta}{1+\gamma}\frac{\rho^{1+\gamma}}{\rho_{0}^{\gamma}} + E_{sym}^{loc}(\rho)\rho\delta^{2}
\nonumber \\
&& + \frac{g_{sur}}{2\rho_{0}}(\nabla\rho)^{2} + \frac{g_{sur}^{iso}}{2\rho_{0}}[\nabla(\rho_{n}-\rho_{p})]^{2},
\end{eqnarray}
where the $\rho_{n}$, $\rho_{p}$ and $\rho=\rho_{n}+\rho_{p}$ are the neutron, proton and total densities, respectively, and the $\delta=(\rho_{n}-\rho_{p})/(\rho_{n}+\rho_{p})$ being the isospin asymmetry. The coefficients $\alpha$, $\beta$, $\gamma$, $g_{sur}$, $g_{sur}^{iso}$ and $\rho_{0}$ are set to be the values of -215.7 MeV, 142.4 MeV, 1.322, 23 MeV fm$^{2}$, -2.7 MeV fm$^{2}$ and 0.16 fm$^{-3}$, respectively. A compression modulus of K=230 MeV for isospin symmetric nuclear matter is produced with these parameters. A Skyrme-type momentum-dependent potential is used in the LQMD model \cite{Fe11}
\begin{eqnarray}
U_{mom}=&& \frac{1}{2\rho_{0}}\sum_{i,j,j\neq i}\sum_{\tau,\tau'}C_{\tau,\tau'}\delta_{\tau,\tau_{i}}\delta_{\tau',\tau_{j}}\int\int\int d \textbf{p}d\textbf{p}'d\textbf{r}   \nonumber \\
&& \times f_{i}(\textbf{r},\textbf{p},t) [\ln(\epsilon(\textbf{p}-\textbf{p}')^{2}+1)]^{2} f_{j}(\textbf{r},\textbf{p}',t).
\end{eqnarray}
Here $C_{\tau,\tau}=C_{mom}(1+x)$, $C_{\tau,\tau'}=C_{mom}(1-x)$ ($\tau\neq\tau'$) and the isospin symbols $\tau$($\tau'$) represent proton or neutron. The parameters $C_{mom}$ and $\epsilon$ was determined by fitting the real part of optical potential as a function of incident energy from the proton-nucleus elastic scattering data. In the calculation, we take the values of 1.76 MeV, 500 c$^{2}$/GeV$^{2}$ for the $C_{mom}$ and $\epsilon$, respectively, which result in the effective mass $m^{\ast}/m$=0.75 in nuclear medium at saturation density for symmetric nuclear matter. The parameter $x$ as the strength of the isospin splitting with the value of -0.65 is taken in this work, which has the mass splitting of $m^{\ast}_{n}>m^{\ast}_{p}$ in nuclear medium. The effect of the momentum dependence of the symmetry potential in heavy-ion collisions was also investigated with the isospin-dependent Boltzmann Uehling Uhlenbeck transport model. The same conclusions are found \cite{Ga11}.

The symmetry energy is composed of three parts, namely the kinetic energy from fermionic motion, the local density-dependent interaction and the momentum-dependent potential as
\begin{equation}
E_{sym}(\rho)=\frac{1}{3}\frac{\hbar^{2}}{2m}\left(\frac{3}{2}\pi^{2}\rho\right)^{2/3}+E_{sym}^{loc}(\rho)+E_{sym}^{mom}(\rho).
\end{equation}
The local part is adjusted to mimic predictions of the symmetry energy calculated by microscopical or phenomenological many-body theories and has two-type forms as follows:
\begin{equation}
E_{sym}^{loc}(\rho)=\frac{1}{2}C_{sym}(\rho/\rho_{0})^{\gamma_{s}},
\end{equation}
and
\begin{equation}
E_{sym}^{loc}(\rho)=a_{sym}(\rho/\rho_{0})+b_{sym}(\rho/\rho_{0})^{2}.
\end{equation}
The parameters $C_{sym}$, $a_{sym}$ and $b_{sym}$ are taken as the values of 52.5 MeV, 43 MeV, -16.75 MeV. The values of $\gamma_{s}$=0.5, 1., 2. have the soft, linear and hard symmetry energy with baryon density, respectively, and the Eq. (7) gives a supersoft symmetry energy, which cover the largely uncertain of nuclear symmetry energy, particularly at supra-saturation densities. All cases cross at saturation density with the value of 31.5 MeV. We chose two typical variations with baryon density, i.e., hard and supersoft symmetry energies in the domain of high densities.

The hyperon mean-field potential is constructed on the basis of the light-quark counting rule. The self energies of hyperons are assumed to be two thirds of that experienced by nucleons. Thus, the in-medium dispersion relation reads
\begin{equation}
\omega(\textbf{p}_{i},\rho_{i})=\sqrt{(m_{H}+\Sigma_{S}^{H})^{2}+\textbf{p}_{i}^{2}} + \Sigma_{V}^{H}
\end{equation}
with $\Sigma_{S}^{H}= 2 \Sigma_{S}^{N}/3$ and $\Sigma_{V}^{H}= 2 \Sigma_{V}^{N}/3$, which leads to the optical potential at the saturation density being the value of -32 MeV. The evolution of mesons (here mainly pions and kaons) is also determined by the Hamiltonian, which is given by
\begin{eqnarray}
H_{M}&& = \sum_{i=1}^{N_{M}}\left( V_{i}^{\textrm{Coul}} + \omega(\textbf{p}_{i},\rho_{i}) \right).
\end{eqnarray}
Here the Coulomb interaction is given by
\begin{equation}
V_{i}^{\textrm{Coul}}=\sum_{j=1}^{N_{B}}\frac{e_{i}e_{j}}{r_{ij}},
\end{equation}
where the $N_{M}$ and $N_{B}$ are the total numbers of mesons and baryons including charged resonances. We consider two scenarios for kaon (antikaon) propagation in nuclear medium, one with and one without medium modification. From the chiral Lagrangian the kaon and antikaon energy in the nuclear medium can be written as \cite{Li97,Ka86}
\begin{equation}
\omega_{K}(\textbf{p}_{i},\rho_{i})=\left[m_{K}^{2}+\textbf{p}_{i}^{2}-a_{K}\rho_{i}^{S}+
(b_{K}\rho_{i})^{2}\right]^{1/2}+b_{K}\rho_{i},
\end{equation}
\begin{equation}
\omega_{\overline{K}}(\textbf{p}_{i},\rho_{i})=\left[m_{\overline{K}}^{2}+\textbf{p}_{i}^{2}-a_{\overline{K}}\rho_{i}^{S}+
(b_{K}\rho_{i})^{2}\right]^{1/2}-b_{K}\rho_{i},
\end{equation}
respectively. Here the $b_{K}=3/(8f_{\pi}^{2})\approx$0.32 GeVfm$^{3}$, the $a_{K}$ and $a_{\overline{K}}$ are 0.18 GeV$^{2}$fm$^{3}$ and 0.3 GeV$^{2}$fm$^{3}$, respectively, which result in the strengths of repulsive kaon-nucleon (KN) potential and of attractive antikaon-nucleon potential with the values of 25.5 MeV and -96.8 MeV at saturation baryon density, respectively. The values of $m^{\ast}_{K}/m_{K}$=1.05 and $m^{\ast}_{\overline{K}}/m_{\overline{K}}$=0.8 at normal baryon density are concluded with the parameters. The effective mass is used to calculate the threshold energy for kaon and antikaon production, e.g., kaon production in the pion-baryon collisions $\sqrt{s_{th}}=m^{\ast}_{Y} + m^{\ast}_{K}$.

The scattering in two-particle collisions is performed by using a Monte Carlo procedure, in which the probability to be a channel in a collision is calculated by its contribution of the channel cross section to the total cross section. The primary products in nucleon-nucleon (NN) collisions in the region of 1\emph{A} GeV energies are the resonances of $\Delta$(1232), $N^{\ast}$(1440), $N^{\ast}$(1535) and the pions. We have included the reaction channels as follows:
\begin{eqnarray}
&& NN \leftrightarrow N\triangle, \quad  NN \leftrightarrow NN^{\ast}, \quad  NN
\leftrightarrow \triangle\triangle,  \nonumber \\
&& \Delta \leftrightarrow N\pi,  N^{\ast} \leftrightarrow N\pi,  NN \leftrightarrow NN\pi (s-state),  \nonumber \\
&& N^{\ast}(1535) \rightarrow N\eta.
\end{eqnarray}
At the considered energies, there are mostly $\Delta$ resonances which disintegrate into a $\pi$ and a nucleon in the evolutions. However, the $N^{\ast}$ yet gives considerable contribution to the energetic pion yields. The energy and momentum-dependent decay widths are used in the model for the resonances of $\Delta$(1232) and $N^{\ast}$(1440) \cite{Fe09}. We have taken a constant width of $\Gamma$=150 MeV for the $N^{\ast}$(1535) decay.

\begin{figure*}
\includegraphics{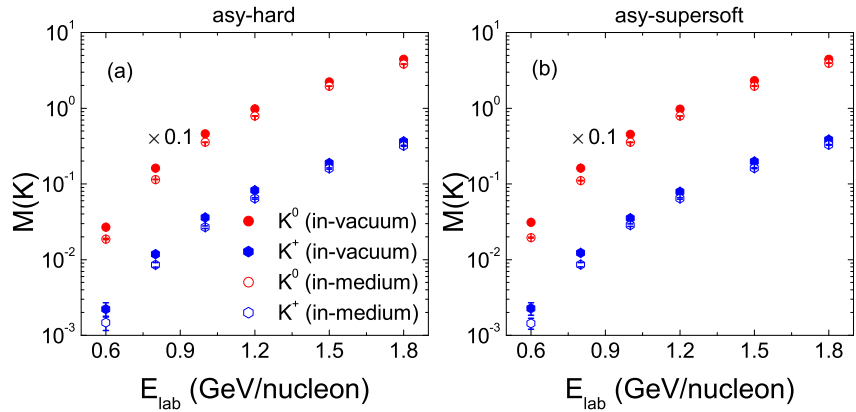}
\caption{\label{fig:wide} (Color online) Total multiplicities of K$^{0}$ and K$^{+}$ as a function of incident energy in central $^{197}$Au+$^{197}$Au collisions at the cases of hard (left panel) and supersoft (right panel) symmetry energies.}
\end{figure*}

The strangeness is created by inelastic hadron-hadron collisions as follows:
\begin{eqnarray}
&& BB \rightarrow BYK,  BB \rightarrow BBK\overline{K},  B\pi \rightarrow YK,  B\pi \rightarrow NK\overline{K}, \nonumber \\
&& Y\pi \rightarrow B\overline{K}, \quad  B\overline{K} \rightarrow Y\pi, \quad YN \rightarrow \overline{K}NN.
\end{eqnarray}
Here the B stands for (N, $\triangle$, N$^{\ast}$) and Y($\Lambda$, $\Sigma$), K(K$^{0}$, K$^{+}$) and $\overline{K}$($\overline{K^{0}}$, K$^{-}$). The elastic scattering between strangeness and baryons are considered through the channels of $KB \rightarrow KB$, $YB \rightarrow YB$ and $\overline{K}B \rightarrow \overline{K}B$. The charge-exchange reactions between the $KN \rightarrow KN$ and $YN \rightarrow YN$ channels are included by using the same cross sections with the elastic scattering, such as $K^{0}p\rightarrow K^{+}n$, $K^{+}n\rightarrow K^{0}p$ etc. Correction of effective mass of kaons in nuclear medium on the elementary cross section is considered through the threshold energy, which results in the reduction of kaon and the enhancement of anti-kaon yields in heavy-ion collisions. Shown in Fig. 1 is a comparison of the total kaon yields produced in the central $^{197}$Au+$^{197}$Au collisions with and without KN potential for the hard (left window) and supersoft (supersoft window) symmetry energies, respectively. The multiplicity of K$^{0}$ is multiplied by 10 times. Inclusion of the KN potential in the model leads to about 30$\%$ reduction of the total kaon yields in the subthreshold domain. Once kaons are produced in the compression stage, the subsequent reabsorption by the surrounding baryons almost does not take place although the distributions of kaons in phase space can be deviated by the in-medium potential.

\begin{figure*}
\includegraphics{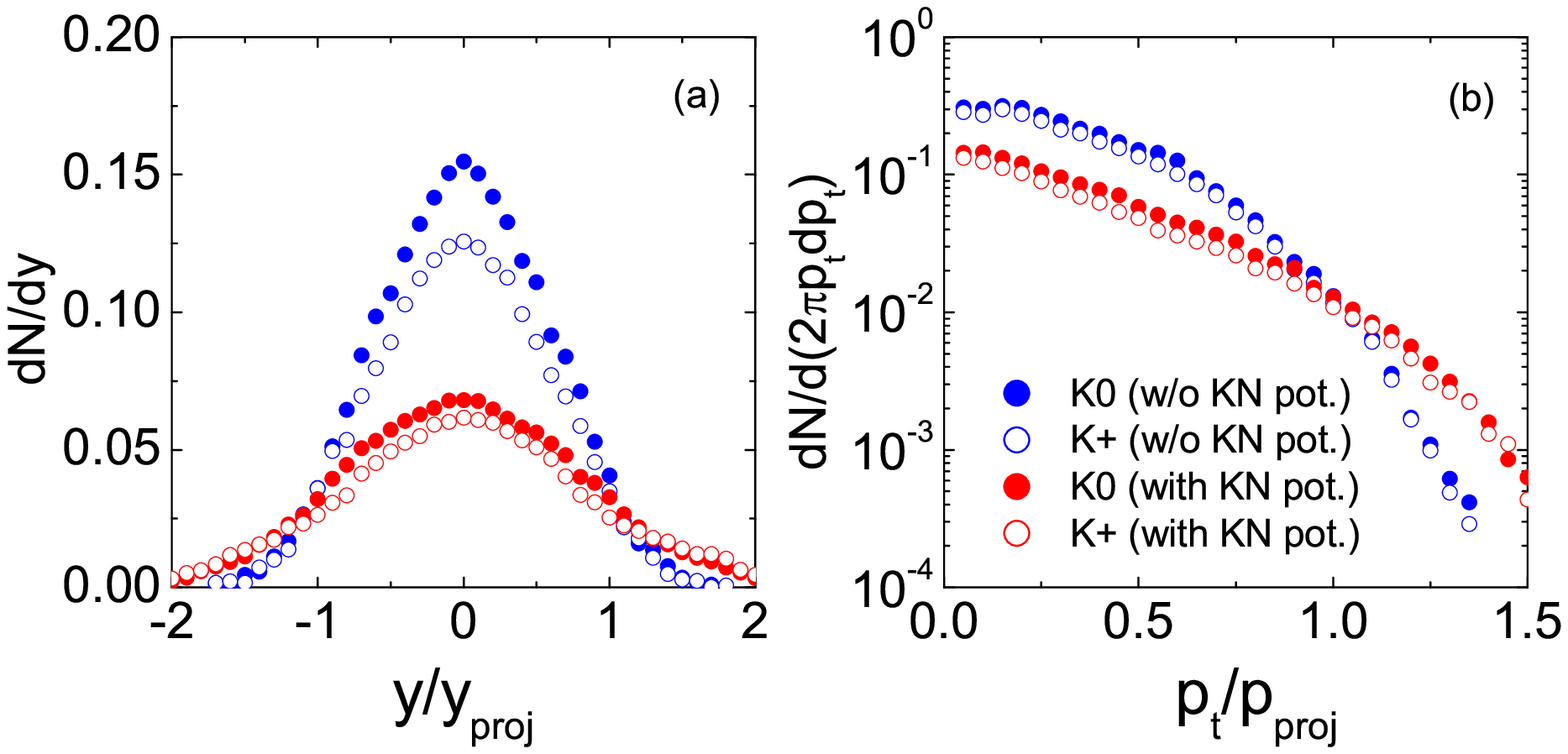}
\caption{\label{fig:wide} (Color online) Longitudinal rapidity (left panel) and transverse momentum (right panel) distributions of isospin kaons in the $^{197}$Au+$^{197}$Au reaction at the incident energy of 1.5\emph{A} GeV.}
\end{figure*}

\begin{figure*}
\includegraphics{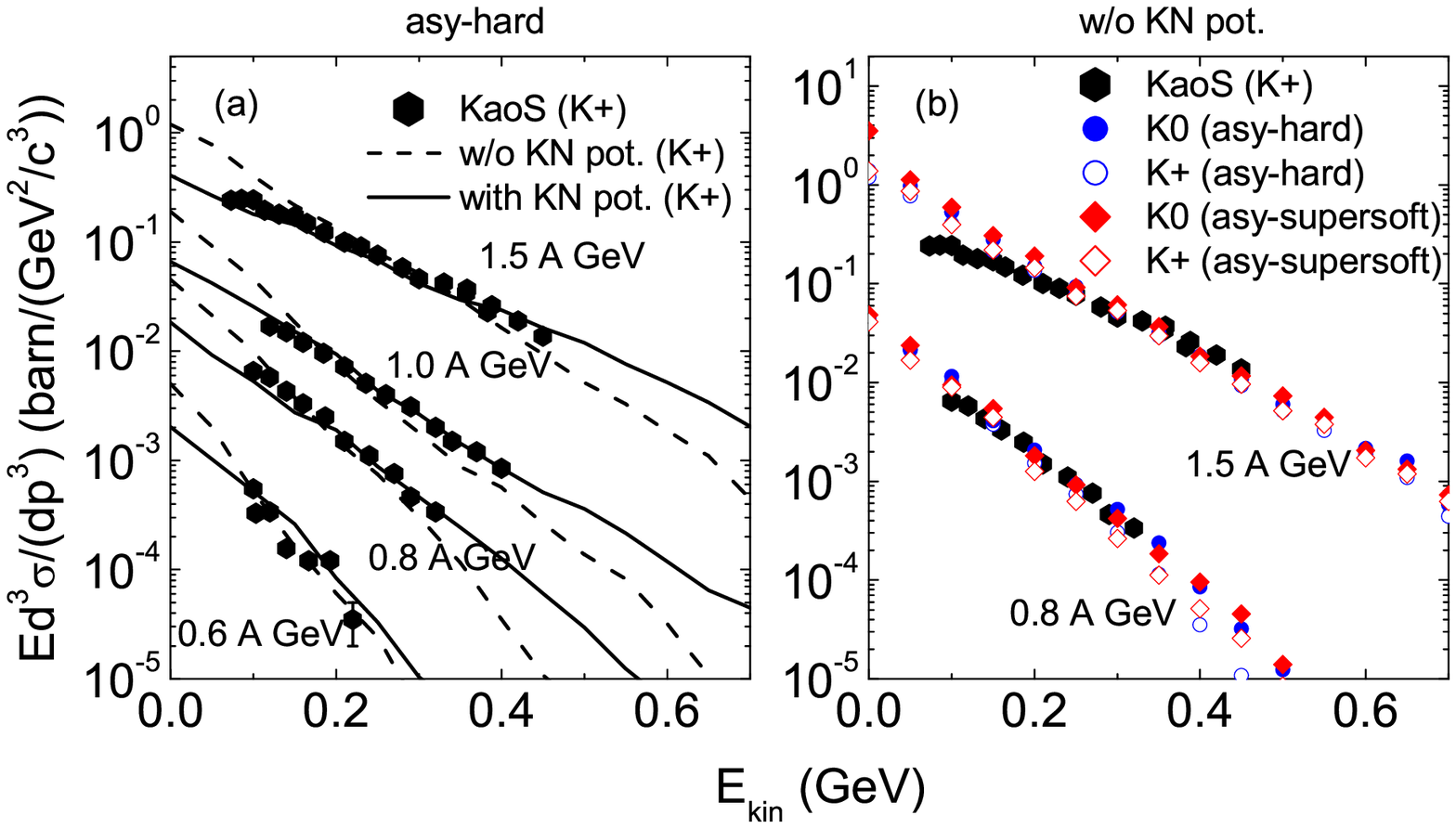}
\caption{\label{fig:wide} (Color online) The kinetic energy spectra of inclusive invariant cross sections in the domain of mid-rapidity with and without inclusion of KN potential (left panel) and for different stiffness of symmetry energies (right panel) in the collisions of $^{197}$Au+$^{197}$Au. The mid-rapidity condition is a selection of $\theta_{c.m.}=90^{\circ}\pm 10^{\circ}$ both for the data and the calculations.}
\end{figure*}

To investigate kaon production in momentum space and its correlation to the collision geometry and to the KN potential, we computed the rapidity and transverse momentum distributions of K$^{0}$ and K$^{+}$ in the near central $^{197}$Au+$^{197}$Au reaction (b=1 fm) with a hard symmetry energy as shown in Fig. 2. One notices that the repulsive KN potential reduces the kaon yields in the mid-rapidity region. A broad rapidity distribution and a flat transverse momentum spectrum are found for the case of the KN potential. It is caused from the fact that the repulsive potential enhances the energetic kaon emission and reduces the kaon yields because of the increase of threshold energy. The isospin effects are pronounced in the domain of mid-rapidity, in particular for the case without inclusion of the KN potential, where kaons are squeezed out from the reaction zone and produced mainly at supra-saturation densities formed during the compression stage of two colliding nuclides. The spectra of inclusive invariant cross sections in the mid-rapidity region for the K$^{+}$ production as measured by the KaoS collaboration \cite{Fo07} in the $^{197}$Au+$^{197}$Au reaction are compared with the LQMD calculations as shown in Fig. 3. A nice agreement between the experimental data and the calculations is obvious with the KN potential. The difference of K$^{0}$ and K$^{+}$ and the influence of nuclear symmetry energy on the spectra can be seen in the domain of higher kinetic energies. Precise measurements on the kaon spectra or its isospin ratio K$^{0}$/K$^{+}$ at high kinetic energy (transverse momentum) are very necessary for constraining the high-density symmetry energy.

\begin{figure*}
\includegraphics{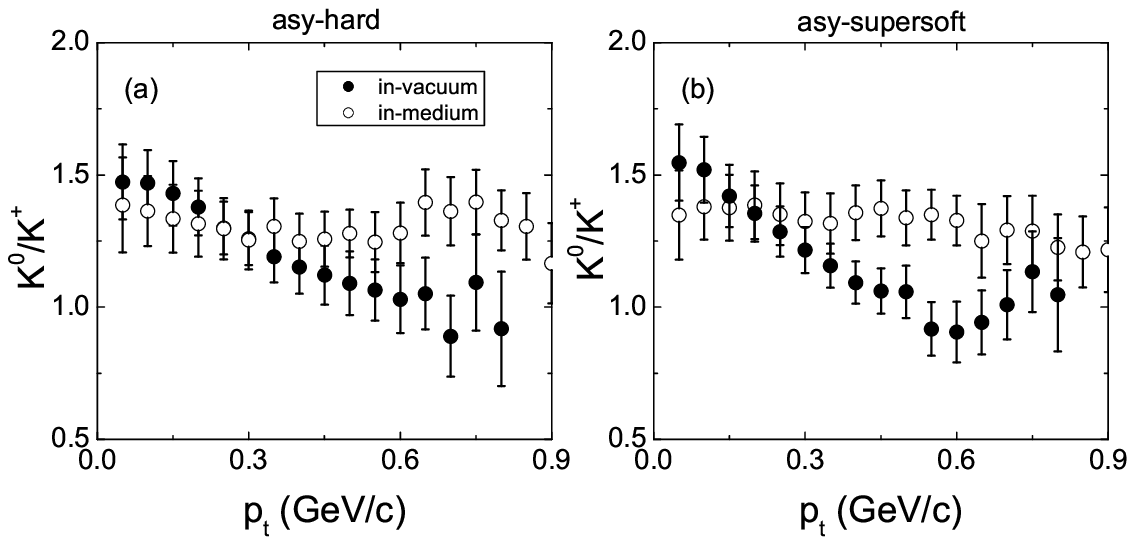}
\caption{\label{fig:wide} Comparison of transverse momentum distributions of the K$^{0}/$K$^{+}$ ratio with and without the KN potentials in central $^{197}$Au+$^{197}$Au collisions at the incident energy of 1.5\emph{A} GeV.}
\end{figure*}

\begin{figure}
\includegraphics[width=8 cm]{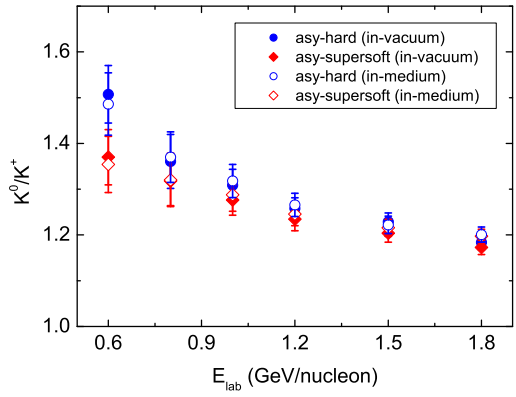}
\caption{\label{fig:epsart} (Color online) Comparison of excitation functions of the $K^{0}/K^{+}$ yields for central $^{197}$Au+$^{197}$Au collisions for the cases of hard and supersoft symmetry energies.}
\end{figure}

\begin{figure}
\includegraphics[width=8 cm]{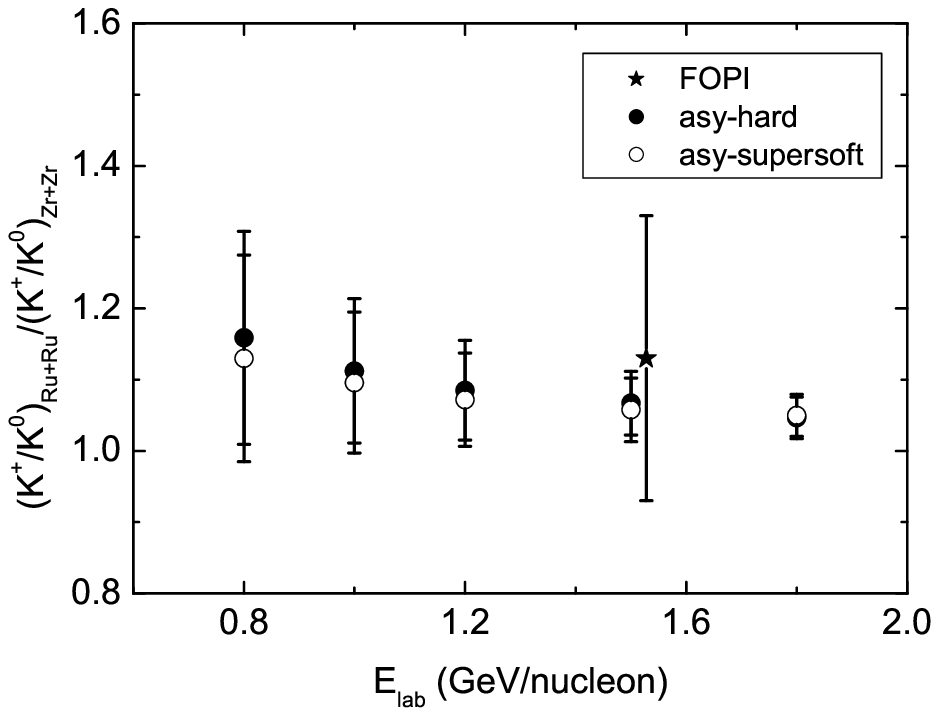}
\caption{\label{fig:epsart} Excitation functions of the double isospin kaon ratios taken from the reactions of $^{96}$Ru+$^{96}$Ru and $^{96}$Zr+$^{96}$Zr in the cases of hard and supersoft symmetry energies and compared with the experimental data \cite{Lo07}.}
\end{figure}

Kaons are produced at the early stage in heavy-ion collision and promptly emitted after production, which can get directly the information of high-density phase diagram \cite{Fe10}. More pronounced effects of the KN potential and the stiffness of symmetry energy on kaon production can be observed from the spectrum of isospin ratio. Shown in Fig. 4 is the influence of the KN potential on the transverse momentum distribution of the K$^{0}/$K$^{+}$ ratio for the hard symmetry energy in the left panel and the supersoft case in the right window. In the both cases the KN potential plays an important role at high transverse momenta and a flat spectrum appears after inclusion of the potential. The results will be helpful for experimental measurements to extract the in-medium kaon potential. The isospin effects appear at deep subthreshold energies as shown in Fig. 5. The in-medium potential slightly changes the K$^{0}/$K$^{+}$ value because of its influence on the kaon propagation and also on the charge-exchange reactions. At the considered energies, the channel of $N\Delta \rightarrow NYK$ contributes the main part for the kaon yields due to the larger production cross sections and the higher invariant energy, and the $NN \rightarrow NYK$ as well as $\pi N \rightarrow YK$ have about one third contributions. One notices that a hard symmetry energy always has the larger values of the isospin ratios than the supersoft case in the domain of subthreshold energies ($E_{th}(K)$=1.58 GeV). This is caused from the enhanced production of $\Delta^{-}$ resonances, i.e., $nn \rightarrow p\Delta^{-}$, $p\Delta^{-} \rightarrow n\Lambda K^{0}$. The threshold effect enlarges the isospin effect of the K$^{0}/$K$^{+}$ yields in the relativistic Boltzmann-Uehling-Uhlenbeck (RBUU) calculations \cite{Fe06}.

A comparison to experimental data is performed from the double ratio of K$^{0}/$K$^{+}$ taken from two systems of $^{96}$Ru+$^{96}$Ru and $^{96}$Zr+$^{96}$Zr as shown in Fig. 6. It should be noted that the influence of symmetry energy on the excitation function of double ratio is very weak over the whole energy range. The less sensitivity to the stiffness of symmetry energy is consistent with the calculations using the RBUU transport model \cite{Lo07}. It is well known that the kaon yields increase with the number of participate nucleons of colliding nuclides. A larger reaction zone with high baryon densities can be formed in heavier colliding systems, where kaons are produced in baryon-baryon and meson-baryon collisions, and the influence of the stiffness of symmetry energy on the isospin kaons is pronounced. Therefore, the isospin ratio in heavy neutron-rich reactions would be a nice probe to extract the density dependence of nuclear symmetry energy.

In summary, kaon dynamics in heavy-ion collisions at near threshold energies has been investigated by using an isospin- and momentum-dependent transport model (LQMD). It is found that the KN potential plays an important role on kaon emission in phase space, in particular, reducing the kaon yields in the mid-rapidity region and also at low transverse momenta, but enhancing the production at high transverse momenta. The isospin ratio of $K^{0}/K^{+}$ depends on the kaon potential at high transverse momenta and a flat spectrum appears after inclusion of the KN potential. The $K^{0}/K^{+}$ ratio of neutron-rich heavy system in the domain of subthreshold energies is sensitive to the stiffness of nuclear symmetry energy, which is a promising probe to extract the high-density information of symmetry energy through comparison to experimental data. Precise measurements on subthreshold kaon production from neutron-rich nuclear collisions are still very necessary.

This work was supported by the National Natural Science Foundation of China under Grant Nos 11175218 and 10805061, the Youth Innovation Promotion Association of Chinese Academy of Sciences, and the Knowledge Innovation Project (KJCX2-EW-N01) of Chinese Academy of Sciences. The author is grateful to the support of K. C. Wong Education Foundation (KCWEF) and DAAD during his research stay in Justus-Liebig-Universit\"{a}t Giessen, Germany.

\end{document}